\documentclass[prl,twocolumn,showpacs,amsmath,amssymb,floatfix,superscriptaddress]{revtex4}
\usepackage{color,graphics,epsfig,rotating}
\usepackage{graphicx}% Include figure files
\usepackage{dcolumn}% Align table columns on decimal point
\usepackage{bm}% bold math
\usepackage{subfigure}

\usepackage{mathrsfs}

\def\beq{\begin{equation}}
\def\eeq{\end{equation}}

\newcommand{\qir}{Q_{\textrm{IR}}}
\newcommand{\qr}{Q_{\textrm{R}}}

\newcommand{\lco}{LCO\,}
\newcommand{\pmo}{PMO\,}
\newcommand{\omr}{\Omega_{\textrm{R}}} %b-coefficient is square of this
\newcommand{\omir}{\Omega_{\textrm{IR}}} %a-coefficient is square of this
\def\ge{g} % Non-linear coupling between the two modes (e-coefficient) 
\newcommand{\ar}{a_3} % Raman mode cubic coefficient (d in PCMO)
\newcommand{\arr}{a_4} % Raman mode quartic coefficient (d in LSMO)
\newcommand{\air}{b_4} % IR mode quartic coefficient (c)

\usepackage[svgnames]{xcolor}
\newif\ifshowcomments\showcommentstrue
\begin{document}

\title{Theory of nonlinear phononics for coherent light-control of solids}

\author{Alaska Subedi}
\affiliation{Centre de Physique Th\'eorique, \'Ecole Polytechnique, CNRS, 91128 Palaiseau Cedex, France}
\author{Andrea Cavalleri}
\affiliation{Max Planck Institute for the Structure and Dynamics of Matter, Hamburg, Germany}
\affiliation{Department of Physics, Oxford University, Clarendon Laboratory, Parks Road, Oxford, UK}
\author{Antoine Georges}
\affiliation{Coll\`ege de France, 11 place Marcelin Berthelot, 75005 Paris, France}
\affiliation{Centre de Physique Th\'eorique, \'Ecole Polytechnique, CNRS, 91128 Palaiseau Cedex, France}
\affiliation{DPMC-MaNEP, Universit\'e de Gen\`eve, CH-1211 Gen\`eve, Switzerland}

\date{\today}

\begin{abstract}
We present a microscopic theory for ultrafast control of solids with
high-intensity Tera-Hertz frequency optical pulses.  
When resonant with selected infrared-active vibrations, these pulses transiently modify the
crystal structure and lead to new collective electronic properties.
The theory predicts the dynamical path taken by the crystal lattice
using first-principles calculations of the energy surface and
classical equations of motion, as well as symmetry considerations.
Two classes of dynamics are identified. 
% In the perturbative regime, Raman-active distortions which preserve crystal symmetry 
%can be induced by cubic anharmonicities. 
%
In the perturbative regime, displacements along the normal mode
coordinate of symmetry-preserving Raman active modes can be achieved
by cubic anharmonicities.
This explains the light-induced insulator-to-metal transition reported
experimentally in manganites. 
We predict a new regime in which %the light can be used to induce
ultra-fast instabilities that break crystal symmetry can be induced.
This non-perturbative effect involves a quartic anharmonic coupling
and occurs above a critical threshold, below which the non-linear
dynamics of the driven mode displays softening and dynamical
stabilization.
\end{abstract}

% \pacs{}

\maketitle

%\section{Introduction}

%% Putting things in context. Light control of specific structural
%% degrees of freedom in solids.  Demonstrated in manganites: Rini et al
%% (MIT), Foerst et al Opens great opportunities for other materials, cf
%% recent work on cuprates etc.  Has been interepreted as non-linear
%% coupling between IR and Raman phonon of $Q^2 Q$ type, leaidng to
%% displacement.

The use of light to control the structural and electronic
properties of solids is emerging as an area of great interest for both
basic research and potential applications. 
Much work has been dedicated to materials with 
strong electronic correlations, which exhibit remarkable collective properties such as superconductivity, 
ferroelectricity or colossal magnetoresistance, 
%and furthermore can be switched between their competing phases by external stimulation, including 
and can be switched by illumination with light
\cite{miya97,cava01,iwai02,chol05,perf06}.

Recently, the possibilities of materials control by light have been
greatly expanded by the demonstration of phase switching through
selective vibrational excitation, that is by deforming the lattice
along a specific normal-mode coordinate \cite{rini07,tobe08,cavi12}.
Let us mention in particular the achievement of light-induced
superconductivity \cite{faus11,kais13,hu14}.
Mode-selective optical control is especially attractive because it allows for the coherent excitation of only one 
or at most a few low-energy degrees of freedom, making control more precise. 
This stands in contrast to what can be achieved at near-visible wavelengths, where the excitation is highly incoherent, 
poorly selective and induces heating. 

A qualitative explanation for mode-selective optical control has been
proposed by F\"{o}rst {\it et al.}\ \cite{foer11}, 
starting from the concepts of ionic Raman scattering \cite{maradudin_wallis_irs_prb_1970,wallis_maradudin_irs_prb_1971,
martin_irs_pss_1974}. In this description, the excitation of an infra-red mode results in the displacement 
of the crystal structure along the coordinate of a second, anharmonically coupled Raman mode. 
The coherent energy transfer from the infra-red to the Raman mode was analyzed through 
a lowest-order perturbative expansion of the lattice potential.
This mechanism was termed ``nonlinear phononics'' and identified as a
new type of coherent coupling between light and crystal lattices,
beyond the extensively studied case of impulsive stimulated Raman
scattering \cite{dhar94,merl97,deko00}.

In this article, we present a microscopic theory for nonlinear phononics. 
This theory is based on symmetry considerations, as well as first-principles %frozen-phonon 
calculations of the energy surface combined with classical equations of motion 
in order to predict the dynamical path taken by the crystal lattice. 
We use this approach to study the response of two materials: the 
parent compound of the magnetoresistive manganites PrMnO$_3$ (\pmo), 
in relation to the experiments of Rini \textit{et al.}\ \cite{rini07}, 
and that of the cuprate superconductors La$_2$CuO$_4$ (\lco). 

At low fluence of the infra-red excitation %, our theory confirms the proposal of F\"{o}rst {\it et al.} 
% In this case, 
we show that, when considering non-degenerate phonon modes, it is only possible to 
displace the crystal lattice along a fully symmetric $A_g$ Raman mode. This makes use of an anharmonic cubic 
coupling of the form $\qir^2\qr$, as envisioned by F\"{o}rst {\it et
  al.}\ \cite{foer11}.
We find that this applies to \pmo and, by performing electronic structure calculations 
based on dynamical mean-field theory (DMFT), we provide a microscopic explanation for the 
light-induced insulator-metal transition observed experimentally\cite{rini07} in this material.
%\cite{rini07}. 

At larger fields, we predict a new class of non-perturbative dynamics that involves Raman distortions of symmetry other than $A_g$.  
This dynamics, which becomes observable in LCO, originates from a quartic coupling of the type $\qir^2\qr^2$. In this 
case, a cubic coupling is forbidden by symmetry. 
Although at low fluence such a quartic coupling implies only a renormalization of the frequency of the Raman mode, an 
instability is found beyond a critical threshold, resulting in a distortion into a crystal structure of lower symmetry.  
%beyond a given threshold the Raman frequency becomes imaginary, resulting in instabilities toward structures of lower symmetry.  
Furthermore, we find that the near-threshold regime exhibits a dynamical stabilization of the crystal lattice, 
analogous to the Kapitza phenomenon in driven non-linear systems \cite{kapitza_1951,stephenson_1908}. 
%
%These predictions could be tested by future experiments on LCO or other materials for which this
%universality class applies. In view of its success for these illustrative cases, we believe that our approach can provide
%theoretical guidance for the selective light-control of materials in a broader context.

\begin{figure} %% [tbp]
  \includegraphics[width=\columnwidth]{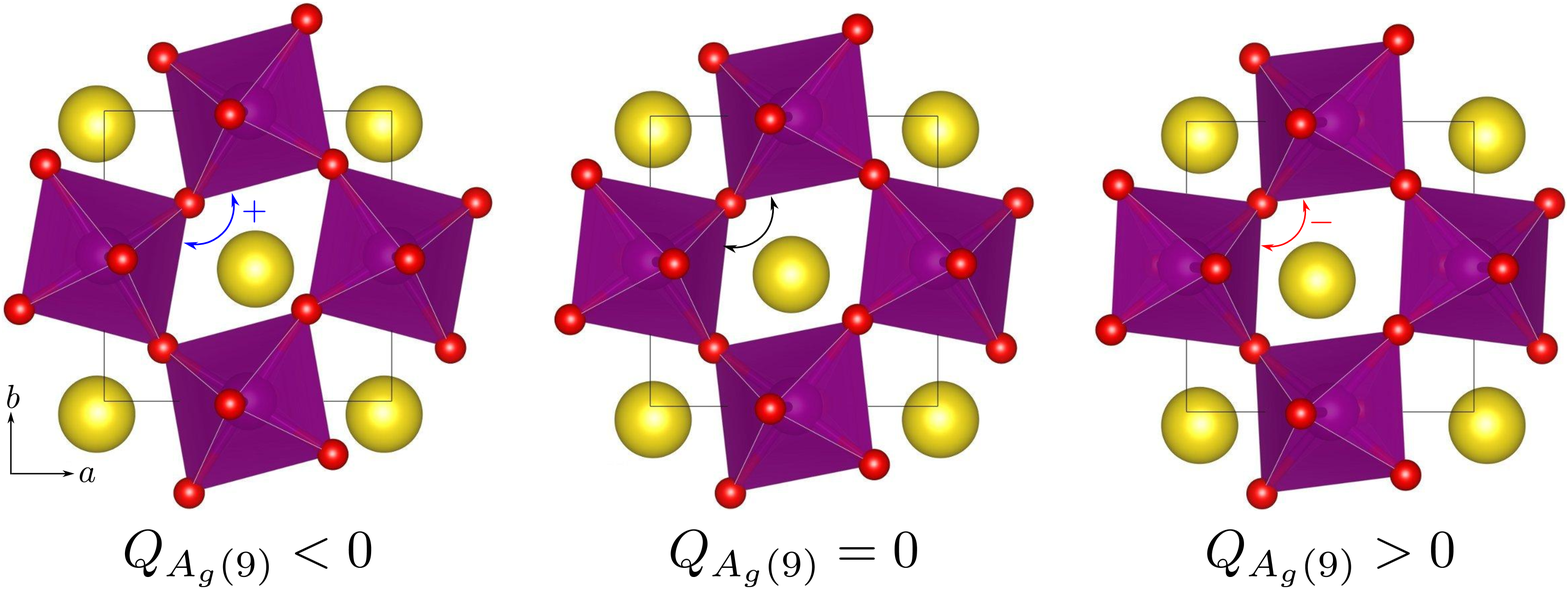}\\
  \includegraphics[width=\columnwidth]{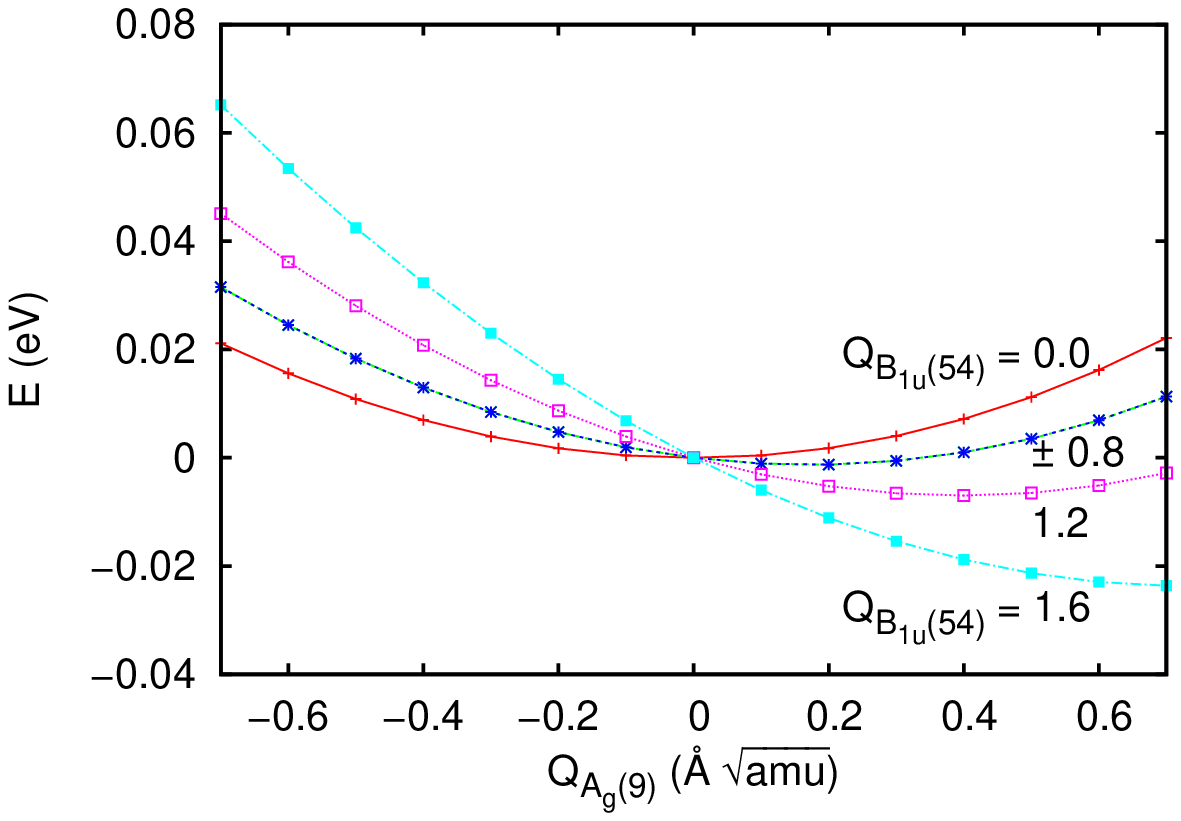}\\
  \includegraphics[width=\columnwidth]{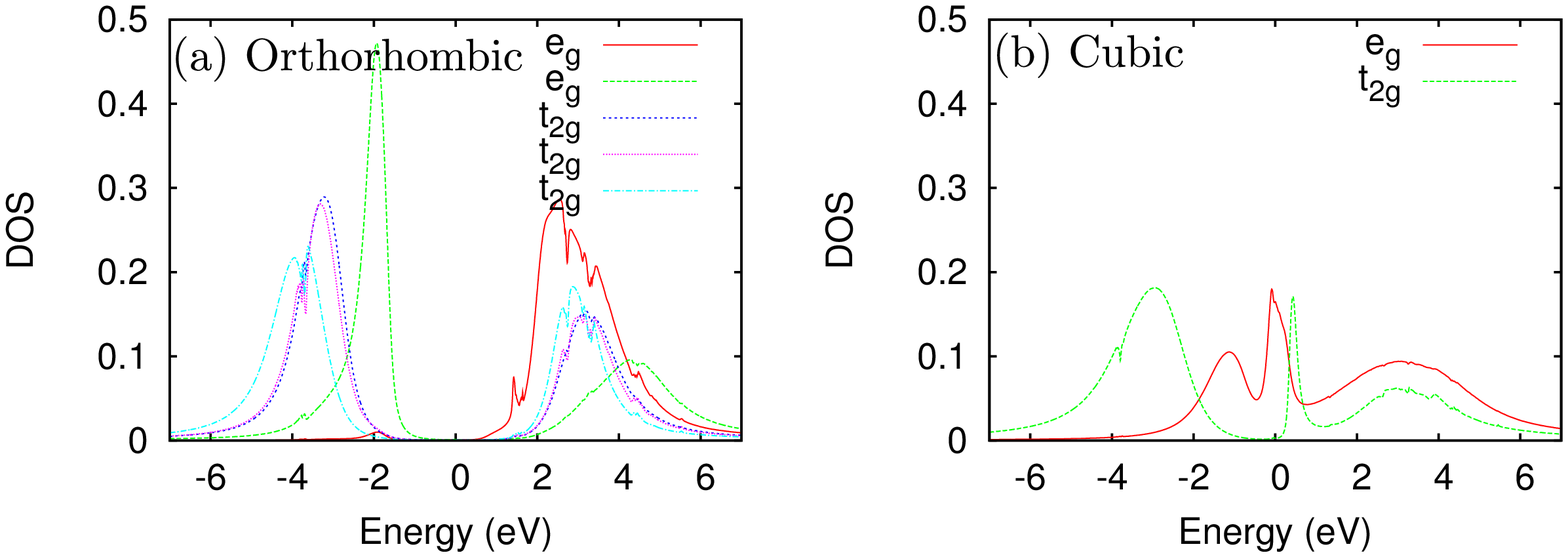}\\
  \caption{
  Top: Sketch of the atomic displacements corresponding to the 
  $A_g(9)$ Raman mode in PrMnO$_3$. A positive displacement drives the angle between octahedra 
  (indicated by the arrow) closer to $90^\circ$, hence the structure closer to the cubic one. 
  Middle: Total energy as a function of the $A_g(9)$ amplitude for
  several values of the $B_{1u}(54)$ amplitude.  For visual clarity,
  we plot $V(\qr,\qir) -V(0,\qir)$ so that all curves coincide at $\qr
  = 0$.
  % Note that the total energy at zero $A_g$-amplitude is conventionally set to zero for each curve. 
%  
  Bottom: LDA+DMFT orbitally-resolved density of states (spectral function) of Mn-$d$ states in \pmo 
  for the equilibrium orthorhombic crystal structure (insulator) and cubic crystal structure (metal). 
    }
  \label{fig:pmo_md54}
\end{figure}

%\section{The two universality classes}
%% We performed energy surface calculations by .etc... ( expalin a bit the method) 
%% for LSMO and PCMO
%% Prseent a bit the materials, symmetry etc...
%%
%% Results displayed in Fig.1
%% As one can see, LSMO has single well at small IR amplitude, a double well is generated
%% beyond a critical IR amplitude. Symmetry always present 
%% 
%% In contrast, PCMO always has a single well, dosplaced from zero as soona s IR mode is vibrating. 
%% No symmetry
%% 
%% This can be understood refer to group theory and to structural drawing in Fig.1
%
%
We first consider PMO, which is an insulator with orthorhombic $Pmna$ structure and four formula units per unit-cell. 
% and is an insulator.  
%
In order to identify the nature and strength of the coupling between various IR and Raman phonons, 
%if they are coupled at all, 
we performed energy-surface calculations as a function of the amplitude of these modes. % for \pmo and \lco. 
The calculations were performed using density-functional theory in a
plane-wave basis set (VASP code \cite{kresse1996b}). 
%
%To obtain the normal modes needed in energy surface calculations, we
%calculated the zone-center modes of PrMnO$_3$ and La$_{0.7}$Sr$_{0.3}$MnO$_3$ 
%using frozen phonon method as implemented in the Phonopy software package.
The 60 zone-center normal modes (see supplemental material for details
\cite{supp}) were identified 
%from zone-center calculations 
using the frozen-phonon method as implemented in the PHONOPY software
package \cite{phonopy}.
%
%PrMnO$_3$ exists in the orthorhombic $Pmna$ structure with four formula units per unit cell and is insulating, 
%while La$_{0.7}$Sr$_{0.3}$MnO$_3$ forms in the rhombohedral $R\overline{3}c$ 
%structure with two formula units per unit cell and is metallic.
%
%It has $60$ zone-center normal modes, as detailed in the supplemental material\cite{supp}. 
%
%La$_{0.7}$Sr$_{0.3}$MnO$_3$ has 30 zone center normal modes with the decomposition \ldots. 
%
In the experiment of Rini \textit{et al.}, the IR mode corresponding to the 
stretching of the apical Mn-O bond was excited, 
but the frequency and symmetry of the excited mode was not analyzed. 
% one cannot directly deduce from the information in Ref.\cite{rini07}
% which Raman mode might have been displaced. 
% but no information about the possible displacement of a Raman mode was provided. 
Hence, we used selection rules from group theory 
%and hints from experiments 
and explored several possible pairs of IR and Raman modes 
to infer which one might be relevant to the physics. 
We considered the excitation of the $B_{1u}(54)$, $B_{1u}(56)$, $B_{2u}(58)$,
and $B_{3u}(60)$ IR modes, the first one corresponding to the stretching of apical 
Mn-O  bonds and the others to in-plane ones. 
In the point group $mmm$, the square of every irreducible representation 
is the $A_g$ representation. Hence, any of the seven $A_g$ modes of PMO can in principle 
have a non-linear cubic coupling to the $B_u$ modes. 
We considered the coupling of all seven modes to 
%seven $A_g$ modes of PrMnO$_3$ to 
the four aforementioned IR modes and found that the coupling between
the apical Mn-O stretching mode $B_{1u}(54)$ and the $A_g(9)$ Raman
mode is substantial.  Furthermore, as illustrated on
Fig.~\ref{fig:pmo_md54} (top panel), a positive amplitude displacement
of this Raman mode reduces the rotation of the MnO$_6$ octahedra in
the $ab$-plane.  We show below that this favours the metallic state.
Hence, we propose that the displacement of this mode through the
non-linear coupling to the pumped $B_{1u}(54)$ mode is responsible for
the effect observed by Rini \textit{et al.}\ \cite{rini07}.

The calculated energy surface is displayed on Fig.~\ref{fig:pmo_md54},
and fits the following expression, which involves a cubic anharmonic
coupling between the two modes:
%
%\begin{eqnarray}
%V(Q_{A_g},Q_{B_{1u}}) & = & a_0 + \frac{1}{2} b Q_{A_g}^2 + \frac{1}{2} a
%Q_{B_{1u}}^2 + \frac{1}{3} d Q_{A_g}^3 + \frac{1}{4} c Q_{B_{1u}}^4
%\nonumber \\ 
%& &-
%\frac{1}{2} e Q_{A_g} Q_{B_{1u}}^2 ,
%\end{eqnarray}
%
\begin{eqnarray}
V(\qr,\qir) & = & \frac{1}{2} \omr^2 \qr^2 + \frac{1}{2} \omir^2\qir^2 
+ \frac{1}{3} \ar \qr^3 + \frac{1}{4} \air \qir^4
\nonumber \\ 
& &-\frac{1}{2} \ge\,\qr\qir^2
\label{eq:energ_pmo}
\end{eqnarray}
%where $Q_{A_g}$ and $Q_{B_{1u}}$ are the displacement of the $A_g(9)$
%and $B_{1u}(54)$ modes, respectively, and the numerical values for the coefficients are summarized in Table I.
%
Here, $\qr$ and $\qir$ are the displacements of the $A_g(9)$ and
$B_{1u}(54)$ modes, respectively. The calculated frequencies are
$\omr=$ 155 cm$^{-1}$ and $\omir=$ 622 cm$^{-1}$.
%
% AS: can't find experimental Ag(9) freq; probably difficult to
% measure because it's too low?
The values of all coefficients are given in a table in the supplemental material \cite{supp}. 
From Fig.~\ref{fig:pmo_md54} (middle panel), one sees that, for a given value of the $B_{1u}(54)$ amplitude,  
the energy landscape has a unique minimum as a function of the amplitude of the $A_g(9)$ mode. 
Furthermore, this mode is displaced from its equilibrium position as soon as the amplitude of the $B_{1u}(54)$ mode is non-zero. 
This displacement being positive, it brings the structure closer to cubic symmetry (Fig.\ref{fig:pmo_md54}, top). 

In order to substantiate that this can be responsible for the observed metallisation of \pmo, 
we have calculated the spectrum of electronic excitations of this compound in both 
the orthorhombic $Pmna$ equilibrium structure and in the hypothetical cubic structure. 
This allows for a clear-cut comparison and for assessing the potential effect of 
fully undoing the orthorombic distortion. 
The calculations are performed using the state-of-the art combination
of electronic structure and dynamical mean-field theory (LDA+DMFT)
\cite{geor96,kotl06,triqs_wien2k_interface,triqs_project} in order to
properly account for the interplay between structural aspects and
strong electronic correlations.
As displayed in Fig.~\ref{fig:pmo_md54} (bottom panel), 
%Fig.\ref{fig:pmo-maxent}, 
we find that \pmo in the $Pmna$ equilibrium structure is an insulator, in agreement with experiments, while the 
hypothetical cubic structure is, remarkably, a metal. 
As detailed in \cite{supp}, the reason for this is the considerable reduction of 
the bandwidth of Mn-$e_g$ states by the orthorhombic distortion, as compared to the
cubic case in which the Mn-O-Mn bonds are straight, leading to a much larger bandwidth. 
As a result, cubic (resp. orthorombic) \pmo are found to be on the metallic (resp. insulating) 
sides of the Mott transition. 
This provides support to our proposal that the excitation of the $A_g$ Raman mode 
by pumping the $B_{1u}$ IR modes explains the insulator to metal transition 
observed by Rini \textit{et al.}\ \cite{rini07}. 

%%%%LCO - Quartic case %%%%%%
%
\begin{figure} %% [tbp]
  \includegraphics[width=\columnwidth]{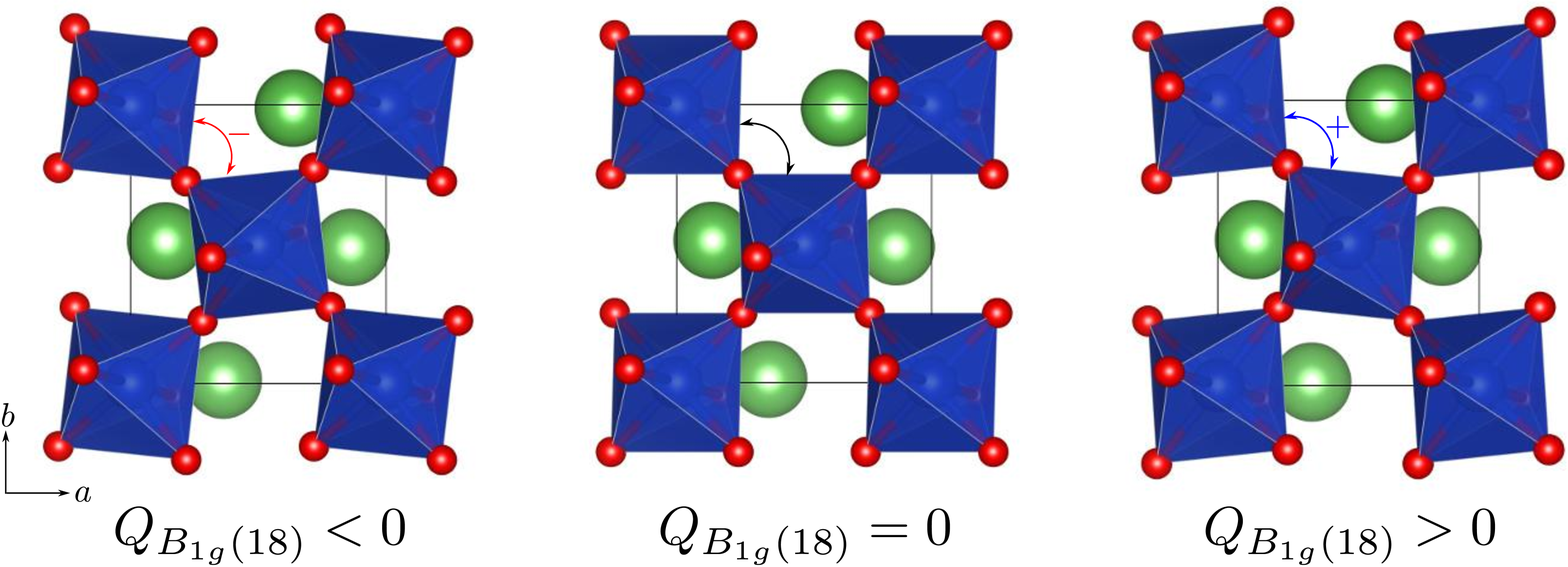}\\
  \includegraphics[width=\columnwidth]{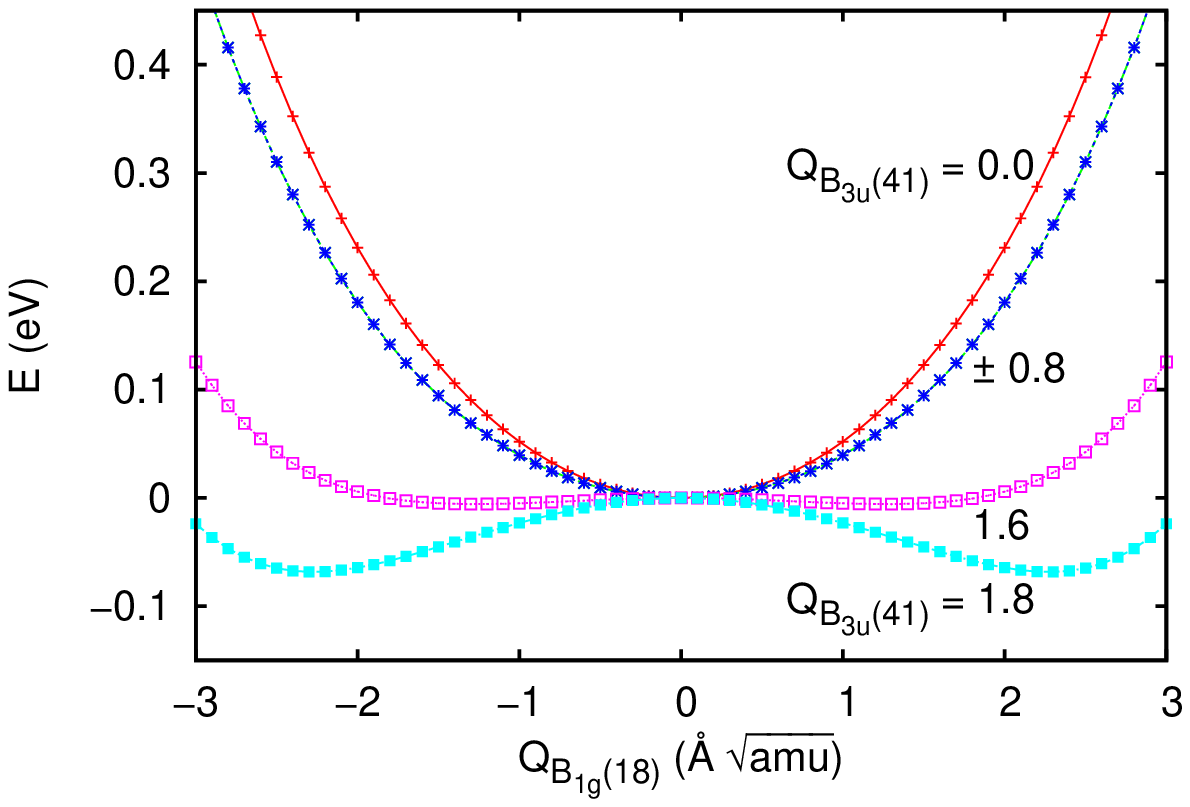}\\
  \caption{Top: Sketch of the atomic displacements corresponding to the $B_{1g}(18)$ Raman mode in La$_2$CuO$_4$, 
  illustrating the symmetry between a negative and positive displacement. 
  Bottom: Total energy %surface 
  as a function of the $B_{1g}(18)$ amplitude for several values of
  the $B_{3u}(41)$ amplitude. 
  For visual clarity, we plot $V(\qr,\qir) -V(0,\qir)$ so that all
  curves coincide at $\qr = 0$.
  % Note that the total energy at zero $B_{1g}$ amplitude is conventionally set to zero for each curve. 
 }
  \label{fig:lco}
\end{figure}
We now turn to \lco, which forms in the orthorhombic $Bmab$ 
structure with two formula units per unit cell. %, and is a metal. 
We looked for coupling between high-frequency IR modes and
low-frequency Raman modes, as relevant to low frequency pump-probe experiments. %on manganites and cuprates. 
%we looked for coupling between high frequency IR modes and low frequency Raman modes in this compound. 
%
We find that there is a substantial coupling between $B_{1g}(18)$
Raman and $B_{3u}(41)$ IR modes of \lco. This IR mode also couples to
a lower frequency $A_g(11)$ Raman mode, but the coupling much
smaller. 
An amplitude of 2.0 \AA$\sqrt{\textrm{amu}}$ for the $B_{3u}(41)$ mode
shifts the minimum of the $A_g(11)$ mode to $-$0.4
\AA$\sqrt{\textrm{amu}}$, whereas the same amplitude of the
$B_{3u}(41)$ mode generates minima at $\pm$2.4
\AA$\sqrt{\textrm{amu}}$ for the $B_{1g}(18)$ mode.
Hence, we only focus on the dynamics of the $B_{1g}(18)$ and
$B_{3u}(41)$ modes. The $B_{1g}(18)$ mode corresponds to the in-plane
rotations of the CuO$_6$ octahedra as shown in Fig.~\ref{fig:lco}
(top) and the $B_{3u}(41)$ involves in-plane stretching of the Cu-O
bonds. The $B_{1g}(18)$ mode breaks the two-fold rotational symmetry
along $x$ and $y$ axes as well as the reflection symmetry with the
mirrors on $xz$ and $yz$ planes. Therefore, the structures generated
by the positive and negative $B_{1g}(18)$ amplitudes are related by
these symmetries.
%
%% The experiment by F\"orst \textit{et al.}\cite{foer11} excites the
%% high frequency Mn-O stretching $E_u$ IR-mode and observes the
%% excitation of the low frequency $E_g$ mode.  We thus consider the
%% coupling between these two modes (among the $30$ zone-center normal
%% modes of this material).
%%
%% In Fig.\ref{fig:lco}, we display the atomic displacements associated with the 
%% $E_g$ mode, which corresponds to rotations of the MnO$_6$ octahedra. 
%% %
%% It is seen that there is a perfect symmetry between a positive and negative 
%% amplitude of this mode, which corresponds to opposite rotation angles. 

The calculated energy surface for \lco is also displayed on
Fig.~\ref{fig:lco} and fits the following expression (with $\qr$ and
$\qir$ the amplitudes of the $B_{1g}(18)$ and $B_{3u}(41)$ modes,
respectively and the calculated frequencies $\omr=$ 162 and $\omir =$
633 cm$^{-1}$):  
%
%\begin{eqnarray}
%V(Q_{E_g},Q_{E_u}) & = & a_0 + \frac{1}{2} b Q_{E_g}^2 + \frac{1}{2} a
%Q_{E_u}^2 + \frac{1}{4} d Q_{E_g}^4 + \frac{1}{4} c Q_{E_u}^4
%\nonumber \\ 
%& &-
%\frac{1}{2} e Q_{E_g}^2 Q_{E_u}^2 ,
%\end{eqnarray}
%
\begin{eqnarray}
V(\qr,\qir) & = & \frac{1}{2} \omr^2 \qr^2 + \frac{1}{2} \omir^2\qir^2 
+ \frac{1}{4} \arr \qr^4 + \frac{1}{4} \air \qir^4
\nonumber \\ 
& &-\frac{1}{2} \ge\,\qr^2\qir^2\,.
\label{eq:energ_lsmo}
\end{eqnarray}
%
%The contrasting energy surface of La$_{0.7}$Sr$_{0.3}$MnO$_3$ as a
%function of the amplitude of the $E_g$ and $E_u$ modes is shown in
%Fig.\ 2.  Unlike the case of PrMnO$_3$, we find
%that the $E_g$ mode of La$_{0.7}$Sr$_{0.3}$MnO$_3$ has a single well
%at small $E_u$ amplitudes, and a double well is generated beyond a
%critical value of the $E_u$ amplitude. Also, the symmetry about the
%equilibrium position is always present in this case.
%
In contrast to \pmo, we find a single-potential well around the
equilibrium value for the $B_{1g}(18)$ mode at small $B_{3u}(41)$
amplitude. Remarkably, a double well is generated beyond a critical
value of the $B_{3u}(41)$ amplitude. Consistent with the observation
above, the energy is symmetric upon reversal of the sign of the
$B_{1g}(18)$ amplitude.  The non-linear coupling is here of the form
$\qr^2\qir^2$, consistent with this symmetry. 
We note that only the coupling to pairs of identical, zone-center
Raman modes have been considered in the analysis of the $\qr^2\qir^2$
coupling term. Coupling to pairs of Raman modes with opposite momenta
away from zone center is in principle possible and should be
considered in future work.

% The following discussion works for centrosymmetric systems. Not sure
% about systems without the inversion symmetry.
%
%The reason for the disparate nature of the potentials of the Raman
%modes in PrMnO$_3$ and La$_{0.7}$Sr$_{0.3}$MnO$_3$ in the presence of an IR mode amplitude 
%can be understood in terms of the symmetry of the phonon modes.  
The disparate nature of the potential energy surface in \pmo and \lco can be explained by symmetry considerations. 
%
%The $A_g$ irrep is the trivial representation. Hence, the
%$A_g$ Raman mode is a fully symmetric mode that does not break any
%symmetry of the crystal. Therefore, the structures that result from
%the atomic displacements according to the positive and negative
%amplitude vibration of equal magnitude of the $A_g$ mode are not
%related by any symmetry, and their total energy can be different. As a
%result, the expansion of the total energy in terms of $Q_{A_g}$ will
%in general have odd powers of $Q_{A_g}$. 
Being associated with the trivial representation of the crystal
symmetry group, an $A_g$ Raman mode does not break any symmetry.
Hence, the atomic displacements associated with vibrations of this
mode whose amplitude have equal magnitude but opposite sign are not
related by symmetry. Their energies are in general different,
resulting in terms with odd powers of $Q_{A_g}$.
In contrast, the $B_{1g}(18)$ mode breaks the rotational and mirror
symmetries mentioned above. 
The structure that results from vibrations of the $B_{1g}(18)$ mode with equal magnitude but opposite signs 
are related by symmetry. They thus have the same energy and only even powers of $Q_{B_{1g}(18)}$ are allowed.  
%a negative and positive amplitude vibration of the $E_g$ mode of an equal magnitude are related by
%symmetry, and their energy must be the same. Therefore, there will
%only be even powers of $Q_{E_g}$ in the expansion of the total energy. 
%
Incidentally, only even powers of the IR mode are allowed in the total energy for similar symmetry reasons. 
%Incidentally, all IR modes also break some symmetries such
%that the negative and positive amplitude structures with same
%magnitude can be transformed to each other. Hence, the IR coordinates
%$\qir$ necessarily come with even powers in the expansion of the total energy.
%
%% It can also be checked that the above form of the coupling between Raman and IR modes 
%% is consistent with the requirement that the total energy should transform as a scalar of the crystal symmetry group. 
%
For non-degenerate phonon modes, it can also be checked that the above
form of the coupling between Raman and IR modes is consistent with the
requirement that the product of the irreducible representations
associated with all phonon modes involved in the coupling should
contain the trivial $A_g$ representation.  Indeed, $A_g \subset A_g
\otimes B_{1u} \otimes B_{1u}$ for \pmo\!, and $A_g \subset B_{1g}
\otimes B_{1g} \otimes B_{3u} \otimes B_{3u}$ for \lco\!. 

% AG: Sentence below commented out: wrong !
%This implies that the product of the irreducible representations associated 
%with all phonon modes involved in the coupling should contain the trivial $A_g$ representation. 
%Indeed, $A_g \subset A_g \otimes B_{1u} \otimes B_{1u}$ for \pmo, and $A_g
%\subset E_g \otimes E_g \otimes E_u \otimes E_u$ for \lsmo. 
%
%The coupling between the Raman and IR modes with the specific irreps
%is also consistent from group theory considerations. The total energy
%is a scalar quantity, so all the terms in its expansion should also
%transform as a scalar. This happens if the trivial irrep $A_g$ is a
%subset of the product of the irreps of the phonon modes in the
%expansion. For PrMnO$_3$, we indeed find that $A_g \subset A_g \otimes
%B_{1u} \otimes B_{1u}$, and for La$_{0.7}$Sr$_{0.3}$MnO$_3$, $A_g
%\subset E_g \otimes E_g \otimes E_u \otimes E_u$.

%\section{Dynamical behaviour from simple model}
%
%% In order to understand dynamics when IR mode is pumped, we resort to
%% simplest posisble model of two coupled classical oscillators.
We finally discuss the dynamics of the non-linearly coupled modes 
%In order to understand the dynamics 
when the IR mode is pumped externally.  
%pumped by the light pulse, 
We simplify the problem by treating the Raman and IR modes as two coupled classical oscillators. 
These oscillators are subject to a force deriving from the calculated energy surface 
(Eqs.~\ref{eq:energ_pmo}, \ref{eq:energ_lsmo}) and to a driving term % due to the light pulse 
$F(t)=F\sin(\Omega t) e^{-t^2/2\sigma^2}$,  
%$F(t)=F\sin(\Omega t) e^{-\frac{t^2}{2\sigma^2}}$,  
%
%We take the expressions obtained from the calculated energy surfaces of Eqs.\ 1 and 2 as the potential
%for the oscillators and derive the equations of motions in the presence of a driving term $F \sin(\Omega t) e^{-\frac{t^2}{2\sigma^2}}$, 
%AG Following sentence can be commented out if need be. 
where $F$, $\sigma$, and $\Omega$ are the amplitude, width and frequency of the light pulse, respectively.
In the case of \pmo (cubic coupling), the resulting equations of motion read: 
\begin{eqnarray}
%\ddot{Q}_{\textrm{IR}}&=& - (\omir^2 - g\qr)\qir -  \air\qir^3 + F(t) \nonumber \\ 
\ddot{Q}_{\textrm{IR}}+\omir^2\qir&=& \ge\,\qr\qir -  \air\qir^3 + F(t) \nonumber \\
%\ddot{\qr}&=& - \omr^2\qr + etc.
\ddot{Q}_{\textrm{R}}+\omr^2\qr&=& \frac{1}{2}\ge\,\qir^2 - \ar \qr^2\,.
\end{eqnarray}
%In the case of PrMnO$_3$, we get the following system of equations of motion:
%\begin{eqnarray}
%\ddot{Q}_{B_{1u}} & = & - (a - e Q_{A_g}) Q_{B_{1u}} - c Q_{B_{1u}}^3
%+ \nonumber \\
% & &  F \sin(\Omega t) e^{-\frac{t^2}{2\sigma^2}} \\ 
%\ddot{Q}_{A_g} & = & - b Q_{A_g} + \frac{1}{2} e Q_{B_{1u}}^2 - d Q_{A_g}^2.
%\end{eqnarray}
%while in the case of \lsmo (quartic coupling) they read:
%\begin{eqnarray}
%\ddot{Q}_{\textrm{IR}}+\omir^2\qir&=& \ge\,\qr^2\qir - \air\qir^3 + F(t) \nonumber \\
%\ddot{Q}_{\textrm{R}}+\omr^2\qr&=& \ge\,\qr\qir^2 - \arr \qr^3
%\end{eqnarray}
Following \cite{foer11}, the resulting dynamical behaviour can be easily understood in the impulsive limit $\sigma\ll 1/\Omega$ 
(see the analysis in supplemental material \cite{supp}, which differs in details from that of \cite{foer11}). 
On resonance, the IR mode undergoes a forced oscillation of amplitude  $\propto F \omir \sigma^{3}$.  
The effective forcing field for the Raman mode is $\ge\qir^2/2\propto gF^2\omir^2\sigma^6 (1-\cos 2\omir t)$ which 
has a rectified non-zero average value. 
Hence, the Raman mode oscillates around a {\it displaced position} 
as a result of the excitation of the IR mode by the light pulse, 
consistent with the displaced minimum of the energy surface (Fig.~\ref{fig:pmo_md54}). 
The displacement occurs for an arbitrarily small pump amplitude, grows quadratically as 
$\propto \ge Q^2_{\textrm{IR,max}}/\omr^2 \propto \ge F^2 \sigma^6\omir^2/\omr^2$, 
and the oscillation frequency ($\omr$) is unchanged by the light pulse. 
This behaviour is also confirmed by numerical integration of these equations for a finite pulse-width \cite{supp}. 
%
%As pointed out in \cite{foer11}, the solution of these equations reveals that 
%the Raman $A_g$ mode oscillates around a {\it displaced position} 
%as a result of the excitation of the IR $B_{1u}$ mode by the light pulse. 
%This can be shown by numerical integration or by an analytical study in the 
%impulsive limit where $\sigma\ll 1/\Omega$. The 
%%
%We integrate this set of equations numerically (see supplemental material\cite{supp}) 
%and find that the Raman $A_g$ mode oscillates around a {\it displaced position} 
%as a result of the excitation of the IR $B_{1u}$ mode by the light pulse. 
%and an example of such vibration for a pulse width
%of 120 fs is shown in Fig.\ 3. Furthermore, the frequency of the Raman
%mode does not change after the pump pulse, although the amplitude of
%the Raman mode varies as the square of the pump amplitude (*check*).
%This system of equations without the anharmonic terms has been studied
%analytically by F\"orst \textit{et al.}, who also find similar
%behavior \cite{foer11}.

\begin{figure} %% [tbp]
  \includegraphics[width=\columnwidth]{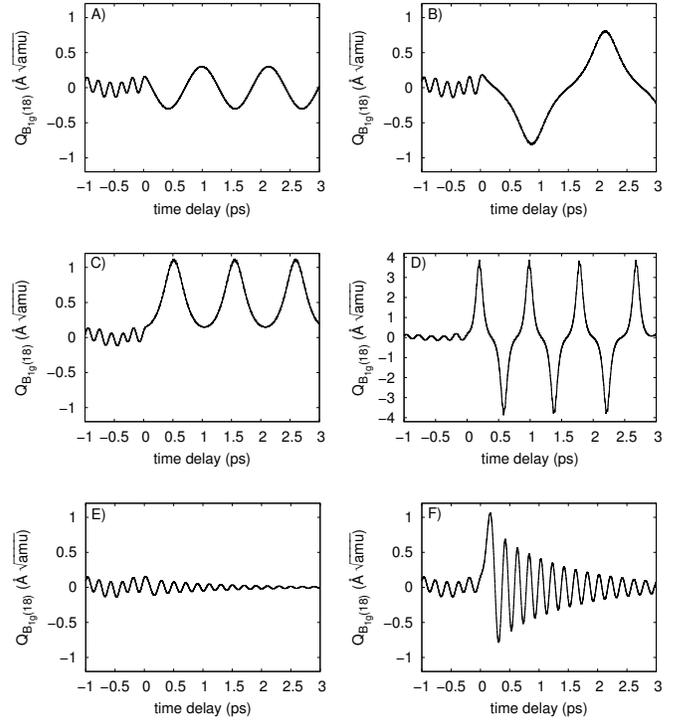}\\
  \caption{Dynamics of the Raman mode in the case of a quartic coupling (as in \lco). 
  Top panel: the four dynamical regimes in the undamped case (A-D). 
  Bottom panel: dynamics below (E) and above threshold (F) in the presence of damping.}
  \label{fig:lco-dyn}
\end{figure}

In the case of \lco (quartic coupling), the equations of motion read: 
\begin{eqnarray}
\ddot{Q}_{\textrm{IR}}+\omir^2\qir&=& \ge\,\qr^2\qir - \air\qir^3 + F(t) \nonumber \\
\ddot{Q}_{\textrm{R}}+\omr^2\qr&=& \ge\,\qr\qir^2 - \arr \qr^3
\label{eq:lsmo_eqs}
\end{eqnarray}
Numerical integration of these equations reveal a much richer behaviour than in the cubic case,  
as illustrated on Fig.~\ref{fig:lco-dyn}. 
There is a threshold value $F_c$ of the pulse amplitude below which the Raman mode 
is not displaced and oscillates around its original equilibrium position. 
In this regime (Fig.\ref{fig:lco-dyn}A), the period of oscillation is amplified 
by the light pulse, as well as its  amplitude.   
Upon increasing $F>F_c$ above threshold, three different regimes are successively found 
(Figs.~\ref{fig:lco-dyn}B,C,D). In a narrow range of $F\gtrsim F_c$, %immediately above threshold, 
a long-period oscillation reaching out to the two wells of the double-well potential is found (B). 
This is also the case (D) at large values of $F$ but with a much shorter period. 
For an intermediate range of values of $F>F_c$, a rectified regime (C) is again found, in which the Raman mode 
oscillates around a displaced position. 

Here also, some analytical  understanding can be achieved %insights can be obtained
in the impulsive limit, as detailed in \cite{supp}. 
It is immediately apparent from the second equation in (\ref{eq:lsmo_eqs}) that the IR mode does not 
act as a forcing term in this case, but rather as a time-dependent modulation of the frequency of the 
Raman mode %$\Omega_{\textrm{eff}}^2(t)=\omr^2 \left[1-\ge\qir^2(t)/\omr^2\right]$. 
$\omr^2\rightarrow\omr^2 \left[1-\ge\qir^2(t)/\omr^2\right]$. 
Hence, for $F<F_c$, the dynamics is well approximated by a Mathieu equation describing a 
parametric oscillator. 
The threshold amplitude $F_c$ can be analytically calculated in the impulsive limit from known 
properties of this equation and is given by the condition 
$\ge Q^2_{\textrm{IR,max}}/\omr^2 = 2$, where $Q_{\textrm{IR,max}}\propto F \omir \sigma^{3}$ 
is the amplitude of the excited IR mode.  
Remarkably, $F_c$ is $\sqrt{2}$ times larger than the value (corresponding to 
$\ge Q^2_{\textrm{IR,max}}/\omr^2 = 1$) at which the energy landscape (\ref{eq:energ_lsmo}) develops a 
double well and the original equilibrium position becomes unstable from a static viewpoint. 
Hence, in the range $F_c/\sqrt{2}<F<F_c$, there is a dynamical stabilisation  of the 
unrectified oscillatory motion, analogous to the Kapitza phenomenon
for a vibrating pendulum \cite{kapitza_1951,stephenson_1908}. 
The oscillation frequency in this regime is reduced by the light pulse according to 
$\Omega_{\textrm{eff}}^2 = \omr^2 \left[1-(F/F_c)^2\right]$, in accordance with the numerical 
observations. %%  \ag{OK ?}. \as{Yes.}

The critical amplitude of the $B_{3u}(41)$ mode associated with the
dynamical threshold for \lco is found to be $Q^c_{\textrm{IR,max}}
\simeq 1.6$ \AA $\sqrt{\textrm{amu}}$. This corresponds to a typical
in-plane displacement of the oxygen atoms of 0.276 \AA.  Previous
experiments \cite{tobe08} have shown that atomic displacements of that
magnitude can be induced by available light pulses, so that this
effect should indeed be observable.
We note furthermore\cite{supp} that the in-plane rotations 
associated with the $B_{1g}(18)$ Raman mode lead to  a reduction of the 
bandwidth of the Cu-$d_{x^2-y^2}$/O-$p_{x,y}$ antibonding band. 
This could be used to modify the correlation strength and superexchange 
coupling in this cuprate. 

%\section{Effect of damping}
%
We have also considered the effect of damping on this dynamical behaviour, as 
displayed on Fig.~\ref{fig:lco-dyn}~E--F. 
For typical values ($5$-$10\%$ of the linewidth)  of the damping 
of the IR mode, the Raman mode is found to 
relax back to an oscillating behaviour around its original equilibrium position, as 
expected.  However, the salient features of the above dynamical regimes 
are preserved. In particular, a critical value of the pump amplitude is still 
present, above which the pump excites a coherent oscillation of the 
Raman mode, with an initial large-amplitude displacement (Fig.\ref{fig:lco-dyn}~F). 

In summary we have developed a microscopic theory of the light-control
of crystal lattices by strong field THz radiation. Our theory greatly
expands the understanding of this new class of phenomena, explaining
many of the observations reported to date. 
%The cubic dynamics, already
The cubic anharmonic coupling between IR and Raman modes, already  
used to explain experimental results qualitatively, is shown here to
apply only to totally symmetric modes (in the non-degenerate case considered here). 
For manganites, we combine first-principle structural studies and 
electronic structure calculations, and identify the microscopic pathway 
for the insulator to metal transition observed experimentally. 
More importantly, we predict a new regime for which the light can be used
to initiate dynamical symmetry breaking. This latter class of
phenomena is non-perturbative, involves a quartic coupling, and leads to 
transient mode softening and dynamic stabilization. 
We show that this second regime is dominant in specific crystal structures, 
and predict that its experimental observation is possible with current technology.

\acknowledgments

We are thankful to Leonid Pourovskii and Michel Ferrero for assistance
with the LDA+DMFT calculations, and to Roman Mankowski,
Michael F\"orst, Roberto Merlin and Jean-Marc Triscone for discussions and suggestions. 
This work was supported by a grant (ERC-319286 QMAC) from the European Research Council 
and by the Swiss National Supercomputing Centre (CSCS) under project ID s404. 

% \bibliography{refs_main}

\newpage

%% % TO DO:
%% % Quote papers on ionic Raman scattering !
%% %
%% \documentclass[prl,twocolumn,showpacs,amsmath,amssymb,floatfix,superscriptaddress]{revtex4}
%% \usepackage{color,graphics,epsfig,rotating}
%% \usepackage{graphicx}% Include figure files
%% \usepackage{dcolumn}% Align table columns on decimal point
%% \usepackage{bm}% bold math
%% % \usepackage[format=plain,justification=centerlast]{caption}
%% \usepackage{subfigure}
%% % \usepackage{subfig}
%% 
%% \usepackage{mathrsfs}

\def\beq{\begin{equation}}
\def\eeq{\end{equation}}

%% \newcommand{\omu}{\Omega_u}
%% \newcommand{\omg}{\Omega_g}
%% \newcommand{\omeff}{\Omega_{\mathrm{eff}}}
%% \newcommand{\xu}{X_u}
%% \newcommand{\xg}{X_g}
%% \newcommand{\xm}{X_m}
%% \newcommand{\ee}{\epsilon}
%% \newcommand{\am}{a_M}
%% \newcommand{\qm}{q_M}

% Added AG
%% \newcommand{\qir}{Q_{\textrm{IR}}}
%% \newcommand{\qr}{Q_{\textrm{R}}}
%% \newcommand{\lsmo}{LSMO\,}
%% \newcommand{\pmo}{PMO\,}
%% \newcommand{\lco}{LCO\,}
%% \newcommand{\omr}{\Omega_{\textrm{R}}} %b-coefficient is square of this
%% \newcommand{\omir}{\Omega_{\textrm{IR}}} %a-coefficient is square of this
%% \def\ge{g} % Non-linear coupling between the two modes (e-coefficient) 
%% %\newcommand{\ar}{a_{3}^\textrm{R}} % Raman mode cubic coefficient (d in PCMO)
%% \newcommand{\ar}{a_3} % Raman mode cubic coefficient (d in PCMO)
%% %\newcommand{\arr}{a_{4}^\textrm{R}} % Raman mode quartic coefficient (d in LSMO)
%% \newcommand{\arr}{a_4} % Raman mode quartic coefficient (d in LSMO)
%% %\newcommand{\air}{a_{4}^\textrm{IR}} % IR mode quartic coefficient (c)
%% \newcommand{\air}{b_4} % IR mode quartic coefficient (c)
%% 
\newcommand{\qirmax}{Q_{\textrm{IR,max}}}

%% \usepackage[svgnames]{xcolor}
%% \newif\ifshowcomments\showcommentstrue
%% %\newif\ifshowcomments\showcommentsfalse
%% \newcommand{\ag}[1]{\ifshowcomments{\color{Red}[\textit{#1}]\color{black}}\else{}\fi}
%% \newcommand{\as}[1]{\ifshowcomments{\color{Blue}[\textit{#1}]\color{black}}\else{}\fi}
%% \newcommand{\ac}[1]{\ifshowcomments{\color{Green}[\textit{#1}]\color{black}}\else{}\fi}

%% \begin{document}

%% \title{Theory of nonlinear phononics for coherent light-control of solids\\
%% - Supplemental Material -}

% \title{Control of Crystal Lattices by Light: a Microscopic Theory of Non-Linear Phononics} 

%% \author{Alaska Subedi}
%% \affiliation{Centre de Physique Th\'eorique, \'Ecole Polytechnique, CNRS, 91128 Palaiseau Cedex, France}
%% \author{Andrea Cavalleri}
%% \affiliation{Max Planck Institute for the Structure and Dynamics of Matter, Hamburg, Germany}
%% \affiliation{Department of Physics, Oxford University, Clarendon Laboratory, Parks Road, Oxford, UK}
%% \author{Antoine Georges}
%% \affiliation{Coll\`ege de France, 11 place Marcelin Berthelot, 75005 Paris, France}
%% \affiliation{Centre de Physique Th\'eorique, \'Ecole Polytechnique, CNRS, 91128 Palaiseau Cedex, France}
%% \affiliation{DPMC-MaNEP, Universit\'e de Gen\`eve, CH-1211 Gen\`eve, Switzerland}

%% \date{\today}
%% \pacs{78.47.J-,31.15.A-,63.20.Ry,71.27.+a,71.30.+h}

%% \maketitle

\section{Supplementary Information}

\section{Phonon calculations for \pmo and \lco}

\subsection{Methods}

The results for the phonon dispersions and the non-linear phonon
coupling calculations presented in the paper were obtained using
density functional theory calculations with plane-wave basis sets and
projector augmented wave pseudopotentials \cite{bloechl994,kresse1999}
as implemented in the VASP software package \cite{kresse1996bsupp}. The
interatomic force constants were calculated using the frozen-phonon
method \cite{parl97} and the PHONOPY software package was used to
calculated the phonon frequencies and normal modes
\cite{phonopysupp}. After the normal modes were identified, total energy
calculations were performed as a function of the IR $\qir$ and Raman
$\qr$ amplitudes to obtain the energy surfaces.
The atomic displacements due to an amplitude $Q_\alpha$ of normal mode $\alpha$ is given by
$U_j = \frac{Q_\alpha}{\sqrt{m_j}} w^\alpha_{j}$, where $U_j$ is the displacement
of the $j$th atom, $m_j$ is the mass of this atom, and $w^\alpha_{j}$ is the
corresponding component of the normal-mode vector. 
Note that $w^\alpha$ is normalized and dimensionless. 

The non-linear coupling between the IR and Raman modes were obtained
by fitting the energy surfaces shown in Figs.~(1) and (2) to the
polynomials Eqs.~(1) and (2), respectively, as given in the main
text. These polynomials fit the respective energy surfaces exactly,
hence there are no approximations in the calculations of the
non-linear couplings, beyond that for the exchange-correlation
functional.

We use the units of eV for energies, amu for masses, and \AA
$\sqrt{\textrm{amu}}$ for the normal mode amplitudes
$Q_{\alpha}$. This means that the frequencies obtained from the fits
of the energy surfaces in Figs.~(1) and (2) to Eqs.~(1) and (2),
respectively, are in the units of $\sqrt{\mathrm{eV} /
  \mathrm{amu}}/\mathrm{\AA}$, and we use a conversion factor of
521.471 cm$^{-1}$/($\sqrt{\mathrm{eV} / \mathrm{amu}}/\mathrm{\AA}$)
to get the frequencies in the familiar units of cm$^{-1}$.

\begin{table}[h!tbp]
  \caption{\label{tab:pmo-freqs} Calculated zone center phonon
    frequencies and the symmetries of selected modes of
    orthorhombic PrMnO$_3$. The mode number is given in the
    parenthesis.}
  \begin{ruledtabular}
    \begin{tabular}{ll}
      Calc. freq. (cm$^{-1}$) & Symmetry \\
      \hline
      97.43   &  $A_g(4)$ \\
      154.80  &  $A_g(9)$ \\
      231.10  &  $A_g(21)$ \\
      267.58  &  $A_g(23)$ \\
      351.07  &  $A_g(36)$ \\
      479.49  &  $A_g(47)$ \\
      552.09  &  $A_g(51)$ \\
      622.12  &  $B_{1u}(54)$ \\
      633.38  &  $B_{1u}(56)$ \\
      639.95  &  $B_{2u}(58)$ \\
      660.54  &  $B_{3u}(60)$ \\
    \end{tabular}
  \end{ruledtabular}
\end{table}

\subsection{PMO}

\pmo exists in the orthorhombic $Pbnm$ structure and is
insulating. The distortions from the ideal perovskite structure
consists of the Jahn-Teller distortion and the rotation of the O
octahedra. The rotation of the O octahedra results from relatively
small radii of Pr$^{3+}$ ion and the subsequent mismatch between Pr-O
and Mn-O bond lengths. This distorted structure has four \pmo formula
units per unit cell. In our phonon calculations, we used the
experimental structure for Pr$_{0.7}$Ca$_{0.3}$MnO$_3$ with $a =
5.426$ \AA, $b = 5.478$ \AA, and $c = 7.679$ \AA\ \cite{jira85}, but
relaxed the atomic positions. The DFT+$U$ calculations were done
within the generalized gradient approximation of Perdew, Burke and
Ernzerhof \cite{pbe}. We used a cut-off of 600 eV for plane-wave
expansion. We also used an on-site Coulomb repulsion $U$ = 5.0 eV and
Hund's exchange $J$ = 0.7 eV for the Mn atom and stabilized the
antiferromagnetic ordering.

There are 20 atoms in the orthorhombic unit cell of \pmo, which gives
rise to 60 zone-center normal modes with the decomposition $7A_g +
7B_{1g} + 5B_{2g} + 5B_{3g} + 8A_u + 8B_{1u} + 10B_{2u} +
10B_{3u}$. The phonon frequencies of the seven $A_g$ and the four IR
modes used in our study is given in Table~\ref{tab:pmo-freqs}. The
coefficients of the polynomial for the energy surface of $A_g(9)$ and
$B_{1u}$ modes are given in Table~\ref{tab:coeffs}.

\begin{table}[h!tbp]
  \caption{\label{tab:lco-freqs} Calculated zone center phonon
    frequencies and the symmetries of selected modes of
    orthorhombic La$_2$CuO$_4$. The mode number is given in the
    parenthesis.}
  \begin{ruledtabular}
    \begin{tabular}{ll}
      Calc. freq. (cm$^{-1}$) & Symmetry \\
      \hline
      68.53   &  $B_{1g}(6)$ \\
      91.46   &  $B_{2g}(7)$ \\
      101.92  &  $B_{3g}(8)$ \\
      125.97  &  $A_{g}(11)$ \\
      144.09  &  $B_{3g}(13)$ \\
      162.16  &  $B_{1g}(18)$ \\
      462.68  &  $B_{2u}(38)$ \\
      632.72  &  $B_{3u}(41)$ \\
      636.18  &  $B_{1u}(42)$ \\
    \end{tabular}
  \end{ruledtabular}
\end{table}

\begin{table}[h!tbp]
  \caption{\label{tab:coeffs} The values of the coefficients of the
    polynomial for energy surfaces of Raman and IR modes obtained from
    fit to the calculated energy surfaces.}
  \begin{ruledtabular}
    \begin{tabular}{lcc}
      & \pmo & \lco \\
      \hline
      $\omr^2$ (meV/amu/\AA$^2$) & 87.09 & 103.55 \\
      $\omir^2$ (meV/amu/\AA$^2$) & 1416.13 & 1462.33 \\
      $\ar$ (meV/amu$^{3/2}$/\AA$^3$) & 5.82 & \\
      $\arr$ (meV/amu$^2$/\AA$^4$) & & 8.36 \\
      $\air$ (meV/amu$^2$/\AA$^4$) & 80.24 & 135.18 \\
      $\ge$ (meV/amu$^{3/2}$/\AA$^3$) & 51.74 & \\
      $\ge$ (meV/amu$^2$/\AA$^4$) & & 46.98 \\
    \end{tabular}
  \end{ruledtabular}
\end{table}

\subsection{\lco}

\lco exists in the orthorhombic $Bmab$ structure. This structure is
derived from the ideal body-centered tetragonal structure by
alternating tilts of the O octahedra along the tetragonal [110] axes
which causes a $(\sqrt{2} \times \sqrt{2} \times 1)$ doubling of the
unit cell. As a result, the primitive unit cell in the orthorhombic
structure contains two formula units. In our calculations of the
phonon frequencies, we used the experimental lattice parameters with
$a$ = 5.3568, $b$ = 5.4058, and $c$ = 13.1432 \AA, but relaxed the
internal coordinates. The DFT calculations were done within the local
density approximation and an energy cut-off of 600 eV was used for the
plane-wave basis.

There are 14 atoms in the unit cell, and this results in 42
zone-center normal modes. These phonon modes have the decomposition
$5A_{g} + 4B_{1g} + 3B_{2g} + 6B_{3g} + 4A_{u} + 8B_{1u} + 7B_{2u} +
5B_{3u}$. We looked for coupling between the three highest frequency
IR modes and the low frequency Raman modes. The frequencies and
symmetries of these modes are given in Table~\ref{tab:lco-freqs}. We
find that $B_{1g}(18)$ mode couples to the $B_{3u}(41)$ and
$B_{1u}(42)$ modes. The $A_g(11)$ mode also couples to the
$B_{2u}(42)$, $B_{3u}(41)$, and $B_{1u}(42)$, but the coupling is much
weaker. Therefore, we do not consider its dynamics here. The coupling
strength of the $B_{1g}(18)$ mode to the $B_{3u}(41)$ and $B_{1u}(42)$
modes is the same. Since either $B_{3u}(41)$ or $B_{1u}(42)$ mode can
be exclusively excited by change in polarization of the pump, we only
consider the coupled dynamics of the $B_{1g}(18)$ and $B_{3u}(41)$
modes here. The coefficients of the polynomial for the energy surface
of $B_{1g}(18)$ and $B_{3u}(41)$ modes are given in
Table~\ref{tab:coeffs}.

\begin{figure} %% [tbp]
  \includegraphics[width=\columnwidth]{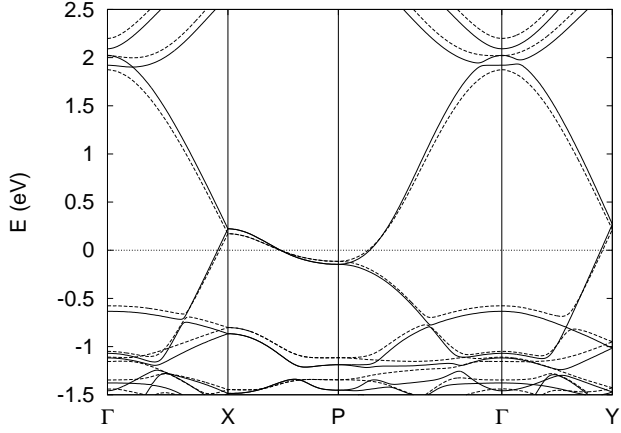}
  \caption{LDA band structures of the orthorhombic La$_2$CuO$_4$ in
    the equilibrium structure (solid lines) and for a structure
    corresponding to $Q_{B_{1g}(18)}=1.5$ \AA $\sqrt{\textrm{amu}}$
    (dashed lines).
  }
  \label{fig:lco-bnd}
\end{figure}

As mentioned in the main text, the $B_{1g}(18)$ mode corresponds to
the in-plane rotations of CuO$_6$ octahedra. Such rotations lead to a reduction of the 
bandwidth of the Cu-$d_{x^2-y^2}$/O-$p_{x,y}$ antibonding band, as
shown in Fig.~\ref{fig:lco-bnd}.

\section{Dynamical behaviour}

\subsection{Excitation of the IR mode}

The dynamics of the IR mode under the action of the pump is described by the equation of motion: 
\beq
\ddot{Q}_{\textrm{IR}}+\omir^2\qir\,=\,F\,\Phi(t)\sin\Omega t\,+\,\cdots
\label{eq:IR_EOM}
\eeq
in which $F$ is a measure of the pump amplitude (with dimension of a force divided by square-root of mass) and 
$\Phi(t)$ is a dimensionless envelope even-function of time describing the shape of the pulse. In the following a Gaussian will 
often be chosen $\Phi(t)=e^{-t^2/2\sigma^2}$ with the characteristic time $\sigma$ a measure of the pulse width.    
Note that the the time-integral of the forcing field $F\Phi(t)\sin(\Omega t)$ integrates to zero over time as 
required by $\int dt E(t)dt = -\int dt \partial A/\partial t =0$. 

The dots $(\cdots)$ in (\ref{eq:IR_EOM}) denote non-linear terms which will be neglected in the present analysis.  
Under this assumption, the general solution of  (\ref{eq:IR_EOM})  reads\cite{landau_mecanique}:
\begin{eqnarray}
\qir(t)\,&=&\,\frac{F}{\omir}\,\int_{-\infty}^{t} d\tau\,\Phi(\tau)\sin(\Omega\tau)\,\sin\left[\omir(t-\tau)\right]\\ \nonumber
&=&\frac{F}{\omir}\sin(\omir t) \int_{-\infty}^{t} d\tau\,\Phi(\tau) \cos(\omir\tau) \sin(\Omega\tau) -\\\nonumber
&-& \frac{F}{\omir}\cos(\omir t) \int_{-\infty}^{t} d\tau\,\Phi(\tau) \sin(\omir\tau) \sin(\Omega\tau)
\end{eqnarray}
We focus on the resonant case $\Omega=\omir$ and consider the impulsive limit $\Omega\sigma \ll 1$. 
The integrals can then be extended to $\tau=+\infty$ and the first one vanishes by parity. In the second integral, one can 
approximate $\sin(\omir\tau)\simeq \omir\tau$, which yields: 
\begin{eqnarray}
\qir(t)\,&=&\,\qirmax\,\cos(\omir t) \label{eq:IR} \\ \nonumber
\qirmax\,&=&\,-F\omir\int_{-\infty}^{+\infty} d\tau \tau^2 \Phi(\tau)
\,=\,-\sqrt{2\pi} F\omir\sigma^3
\end{eqnarray}

\subsection{Dynamics of the Raman mode - cubic case (PMO)}

Neglecting all non-linear terms except the cubic coupling between the IR and Raman mode, the 
equation of motion reads: 
\beq
\ddot{Q}_{\textrm{R}}+\omr^2\qr\,=\,\frac{1}{2}\ge\,\qir(t)^2 + \cdots
\eeq
Using (\ref{eq:IR}), this reads: 
\beq
\ddot{Q}_{\textrm{R}}+\omr^2\qr\,=\,\frac{1}{2}\ge\,\qirmax^2\,\cos^2(\omir t)
\eeq
Rewriting $\cos^2\omir t = (1+\cos 2\omir t)/2$, one sees that the Raman mode undergoes a finite 
displacement: 
\begin{eqnarray}
\delta\qr\,&=&\,\frac{\ge}{4\omr^2}\qirmax^2 \label{eq:displacement} \\ \nonumber
&=& \frac{\ge}{4} F^2 \frac{\omir^2}{\omr^2} \left[\int_{-\infty}^{+\infty} d\tau \tau^2 \Phi(\tau)\right]^2  
\,=\,\frac{\pi}{2}\frac{\omir^2}{\omr^2} \ge F^2\sigma^6
\end{eqnarray}
The Raman mode oscillates at a frequency $\omr$ around this displaced position. This conclusion 
coincides qualitatively with that of Ref.\cite{foer11supp}, although the details of the expressions obtained 
here are different. 

\begin{figure} %% [tbp]
  \includegraphics[width=\columnwidth]{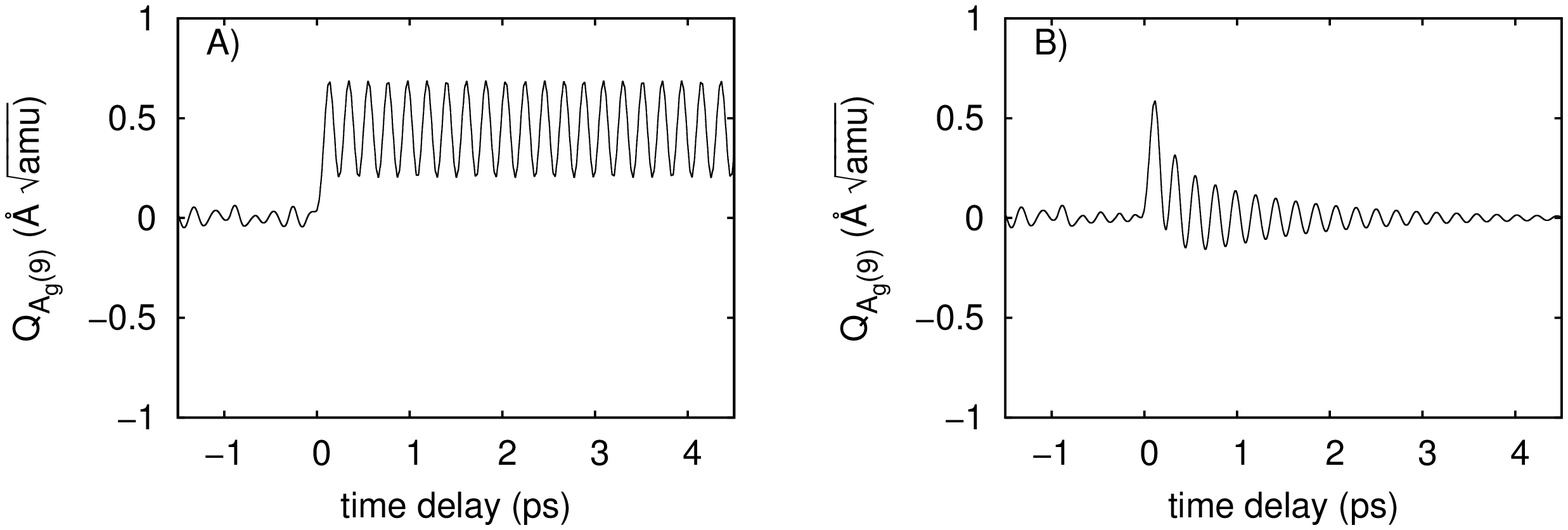}\\
  \caption{Dynamics of the Raman $A_g(9)$ mode for \pmo (cubic coupling). 
  Left panel: dynamics without damping.
  Right panel: dynamics with damping values of 5\% for both
  $B_{1u}(54)$ and $A_g(9)$ modes.}
  \label{fig:pmo-dyn}
\end{figure}

In Fig.\ref{fig:pmo-dyn}, we display the result of a numerical integration of the coupled equations 
for the IR and Raman mode in the case of PMO, which clearly displays the displacement of the Raman mode. 
The second panel of this figure also demonstrates that the displacement still holds in the presence of damping 
(of course, in this case, the mode eventually relaxes back to its original position).  

An alternative method for analysing the dynamics of the Raman mode is to construct 
an {\it effective potential} for this mode by time-averaging over the fast dynamics of the 
IR mode. This is valid because $\omr\ll\omir$, so that the dynamics of the Raman mode 
is much slower.  This method also allows to take into account the non-linearities in the dynamics of 
the Raman mode. Using the time-average:
$\overline{\qir^2(t)}=\qirmax^2\overline{\cos^2(\omir t)}=\qirmax^2/2$, one obtains:
\begin{equation}
V_{\mathrm{eff}}(\qr)=\frac{1}{2}\omr^2\qr^2 +\frac{1}{3}\ar\qr^3 - \frac{1}{4}\ge\qirmax^2\qr
\label{eq:effpot_cubic}
\end{equation}
The displaced position $\delta\qr$ corresponds to the minimum of this potential 
given by $\ar\delta\qr^2+\omr^2\delta\qr-\ge\qirmax^2/4=0$, and thus reads 
\begin{eqnarray}
\delta\qr&=&\frac{\omr^2}{2\ar}\left[\sqrt{1+\frac{\ar\ge\qirmax^2}{\omr^4}}-1 \right]\\ \nonumber
&\simeq& \frac{\ge}{4\omr^2}\qirmax^2 -\frac{1}{16} \ar \frac{\ge^2\qirmax^4}{\omr^6}+\cdots
\end{eqnarray}
The last expression holds for $\ar\ge\qirmax^2/\omr^4\ll 1$, and coincides with (\ref{eq:displacement}) above, plus corrections 
from the non-linear terms. 
We note that the displaced position does not coincide with the minimum of the 
static potential calculated at $\qir=\qirmax$, but rather with that of the effective 
time-averaged potential - the resulting estimate of the displacement is smaller by a factor of two. 

The oscillation frequency at the displaced position is essentially unchanged and given by: 
\begin{eqnarray}
\Omega_{R,\mathrm{eff}}^2&=&\frac{\partial^2 V_{\mathrm{eff}}}{\partial \qr^2}|_{\qr=\delta\qr} 
= \omr^2 + 2\ar\delta\qr  
\\ \nonumber
&=&\omr^2 \left[1 -\frac{1}{8}\left(\frac{\ar\ge\qirmax^2}{\omr^4}\right)^2+\cdots\right]
\end{eqnarray}

\subsection{Dynamics of the Raman mode - quartic case (LCO)}

When the coupling is quartic, the equation of motion of the Raman mode reads: 
\beq
\ddot{Q}_{\textrm{R}}+\omr^2\qr\,=\,\ge\qir^2(t)\, \qr - \arr\qr^3
\eeq
which can be rewritten using (\ref{eq:IR}):
\beq
\ddot{Q}_{\textrm{R}}+\omr^2\left[1-\frac{\ge\qirmax^2}{\omr^2}\cos^2(\omir t)\right]\,\qr\,=\,-\arr\qr^3
\eeq
Neglecting the non-linear term in the r.h.s, this equation describes a parametric oscillator with a frequency which is modulated over 
the fast time-scale corresponding to the period of the IR mode. 
In the limit where $\omir\gg\omr$, the condition for the stability of the oscillatory motion around 
the undisplaced position $\qr=0$ can be derived from a simple physical consideration. Indeed, averaging over the 
fast motion of the IR mode, the effective potential for the Raman mode reads in this case: 
\begin{equation}
V_{\mathrm{eff}}(\qr)\,=\,
%\frac{1}{2}\omr^2\qr^2 \left[1-\frac{\ge\qirmax^2}{\omr^2}\overline{\cos^2(\omir t)}\right] \nonumber \\
\frac{1}{2}\omr^2\qr^2 \left[1-\frac{\ge\qirmax^2}{2\omr^2}\right] + \frac{1}{4}\arr\qr^4
\label{eq:effpot}
\end{equation}
The motion becomes unstable when this effective potential acquires a negative curvature, so that 
to first approximation (ie neglecting corrections of order $\omr/\omir \ll 1$, see below) the 
instability threshold is given by:
\begin{eqnarray}
&&\frac{\ge\qirmax^2}{2\omr^2}=1 \Rightarrow \\ \nonumber 
&&\Rightarrow F_c = \sqrt{\frac{2}{\ge}} \frac{\omr}{\omir} \left[\int_{-\infty}^{+\infty} d\tau \tau^2 \Phi(\tau)\right]^{-1}
= \frac{\omr}{\omir} \frac{1}{\sqrt{\pi\ge}\sigma^3}
\end{eqnarray}
Restoring the non-linear term $\arr\qr^4/4$, the effective (time-averaged) potential develops a double well for $F>F_c$. 
Note that for $F\in [F_c/\sqrt{2},F_c]$ ($\ge\qirmax^2/\omr^2\in[1,2]$), the {\it instantaneous} potential seen by the parametric oscillator 
has negative curvature around $\qr=0$ (and hence a double-well shape) for part of the period. Nonetheless, the 
undisplaced motion is stable in this regime: this is analogous to the Kapitza-Stephenson\cite{kapitza_1951supp,stephenson_1908supp} stabilization of a nominally unstable motion by a 
fast driving force.
Rewriting the equation of motion as (neglecting the non-linear term): 
\beq
\ddot{Q}_{\textrm{R}}+\omr^2\left[1-\frac{\ge\qirmax^2}{2\omr^2}\right]\qr -\ge\qirmax^2\cos(2\omir t)\,\qr\,=\,0
\label{eq:raman_param_lin}
\eeq
we see that the frequency of the Raman mode is renormalized for $F<F_c$ according to: 
\beq
\Omega_{R,\mathrm{eff}}\,=\,\omr\,\left[1-\frac{\ge\qirmax^2}{2\omr^2}\right]^{1/2} = \omr\,\sqrt{1-\frac{F^2}{F_c^2}}
\eeq
Close to threshold, the effective frequency vanishes and the actual behaviour of the amplitude depends 
of course on the non-linearity $\arr$ neglected above (if $\arr$ is neglected, the  amplitude formally diverges at $F_c$). 

A more precise determination of the critical threshold can be obtained by noting that (\ref{eq:raman_param_lin}) is 
actually a Mathieu equation, of the form:
\beq
\frac{d^2y}{dv^2} + \left[a-2q \cos(2v)\right] y\,=\,0
\eeq
in which we have used the standard notations of Abramowitz and Stegun\cite{abramo} with:
\begin{eqnarray}
&&y\equiv \qr\,\,,\,\,
v\equiv \omir t\,\,,\,\,\\ \nonumber
&&a\equiv \frac{\omr^2}{\omir^2}\left[1-\frac{\ge\qirmax^2}{2\omr^2}\right]\,\,,\,\,
q\equiv \frac{\ge\qirmax^2}{4\omir^2}
\end{eqnarray}
As the dimensionless control parameter $\lambda\equiv\frac{\ge\qirmax^2}{2\omr^2}$ is increased, $q$ also 
increases. The threshold is not exactly located at $\lambda=1$ but in fact corresponds to the crossing of the 
separatrix associated with the first characteristic value of the Mathieu equations, given as a power-series in $q\ll 1$ by the equation:
\beq
a\,=\,-\frac{1}{2} q^2 + \frac{7}{128} q^4 +\cdots
\eeq
Using the above expressions of the coefficients $a$ and $q$, this leads to the improved estimate of the threshold, 
including corrections of order $\omr^2/\omir^2$: 
\beq
\frac{\ge\qirmax^2}{2\omr^2}|_c \,=\, 1+ \frac{1}{8} \frac{\omr^2}{\omir^2} + \cdots
\eeq
where the dots stand for higher corrections in $\omr^2/\omir^2\ll 1$.

\section{Details on electronic structure and DMFT calculations - PMO}

\begin{figure} %% [tbp]
  \includegraphics[width=\columnwidth]{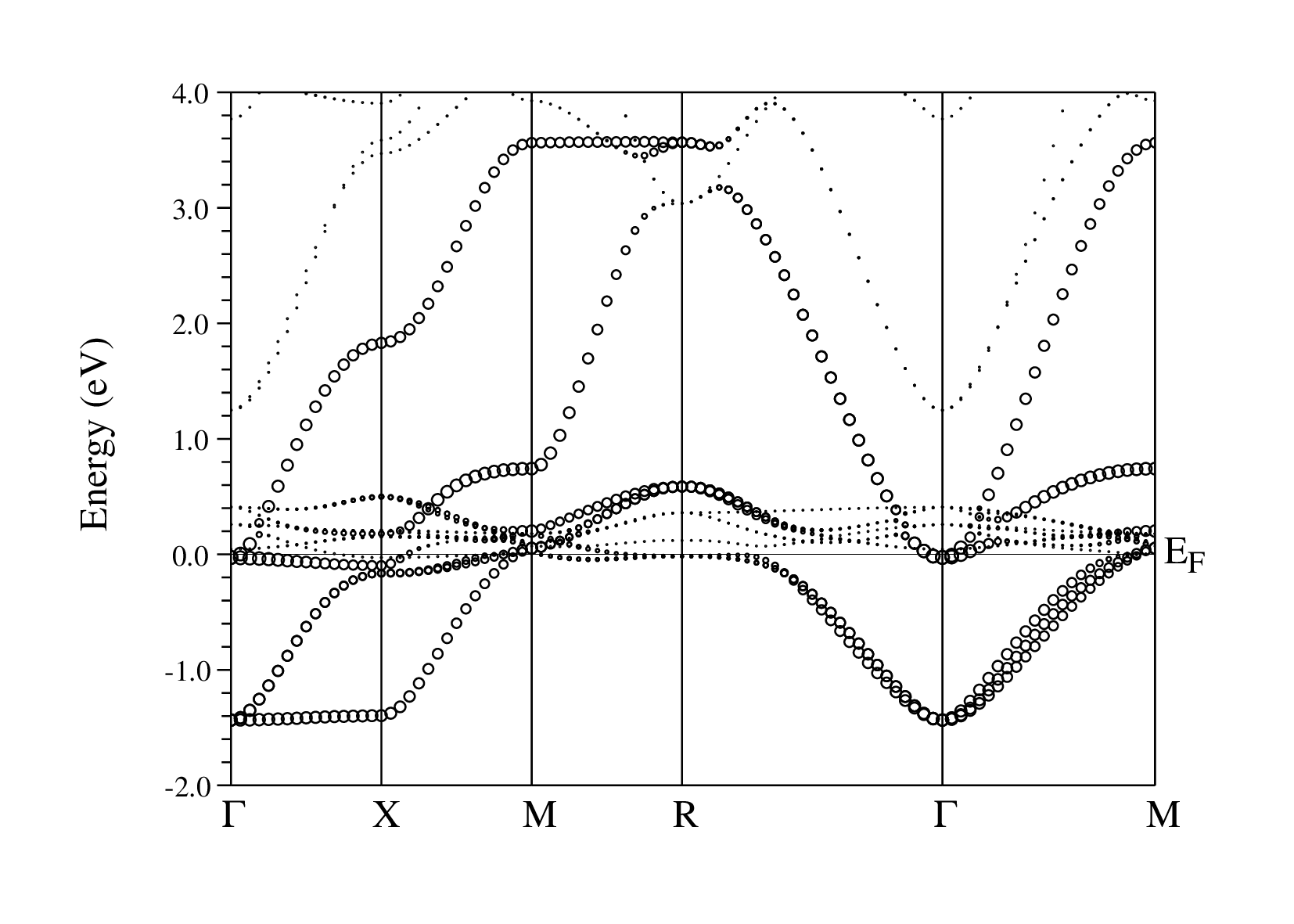}\\
  \includegraphics[width=\columnwidth]{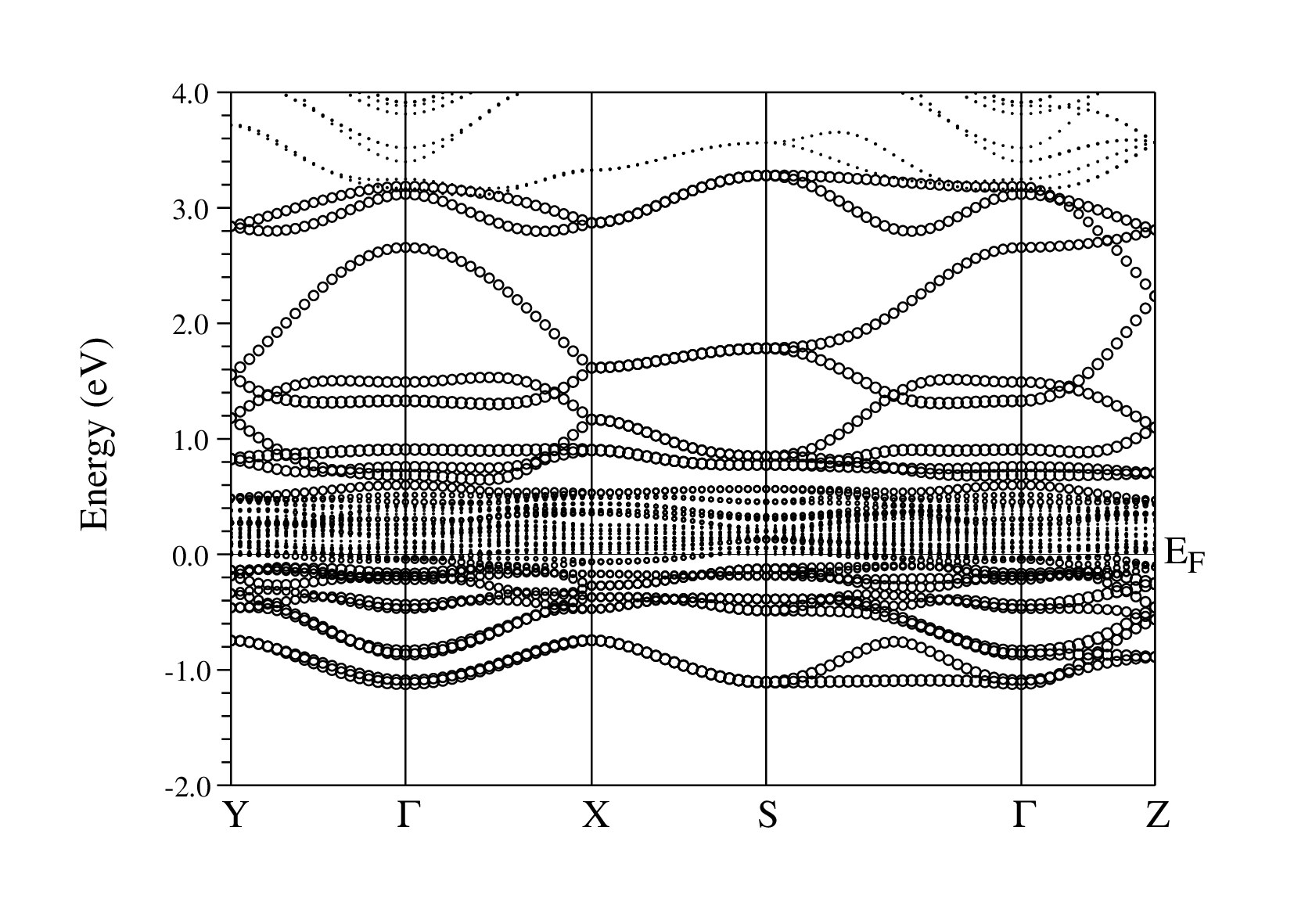}\\
  \caption{(Top) LDA band structure of hypothetical cubic
    \pmo. (Bottom) LDA band structure of orthorhombic \pmo.
    These are plotted with symbols of size proportional to Mn $3d$
    character (plus a very small size to show the position of bands
    with no Mn $3d$ character).
  }
  \label{fig:pmo-bnd}
\end{figure}

The DFT+DMFT calculations on \pmo were performed using a combination
of an all-electron full-potential electronic structure method as
implemented in WIEN2k package \cite{triqs_wien2k_interfacesupp,wien2k} and
DMFT treatment for the Pr $4f$ and Mn $3d$ states as implemented in
TRIQS package \cite{triqs_projectsupp}. The DFT calculations were done
within the local density approximation.  We performed calculations on
the experimental orthorhombic structure\cite{alon00} as well as a
hypothetical cubic structure with the same per formula unit
volume. The band structure calculations show that both the Pr $4f$ and
Mn $3d$ states lie around the Fermi level. Hence, it was necessary to
treat both the Pr $4f$ and Mn $3d$ states using DMFT. We used the
Hubbard-I approximation for the Pr $4f$ states and the numerically
exact hybridization-expansion continuous time quantum Monte-Carlo
\cite{ctqmc,triqs_projectsupp} for the Mn $3d$ states. The Wannier
orbitals were constructed using the projection scheme of
Ref.~\onlinecite{triqs_wien2k_interface}. We used the energy windows
of [-2.0, 4] eV for the cubic structure and [-2.0, 3.4] eV for the
orthorhombic structure. All seven Pr $4f$ and five Mn $3d$ orbitals
lie within this window. We used $U$ = 6.0 and $J$ = 0.7 eV for Pr $4f$
orbitals and $U$ = 5.0 and $J$ = 0.75 eV for Mn $3d$ orbitals. The
calculations were done at an inverse temperature of $\beta = 80$
eV$^{-1}$ ($kT\simeq 150$~K).

\begin{figure} %% [tbp]
  \includegraphics[width=\columnwidth]{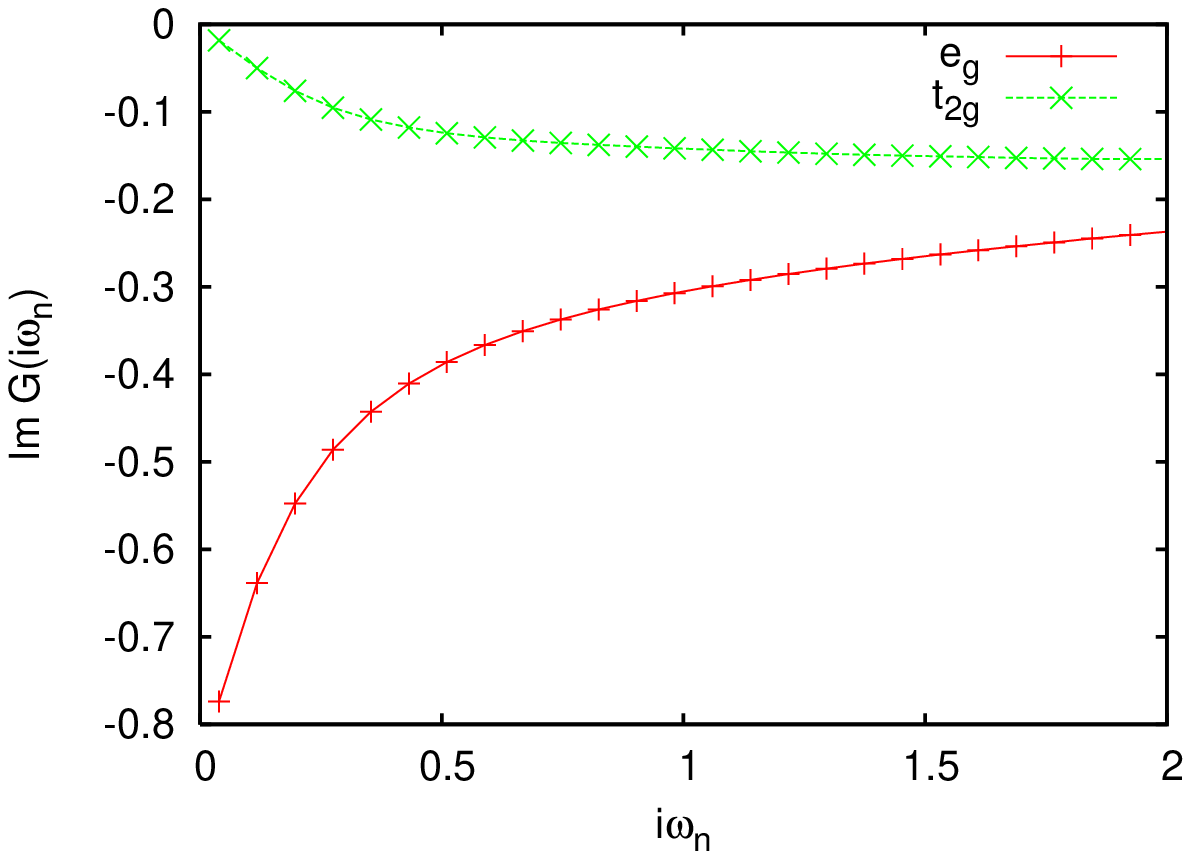}\\
  \includegraphics[width=\columnwidth]{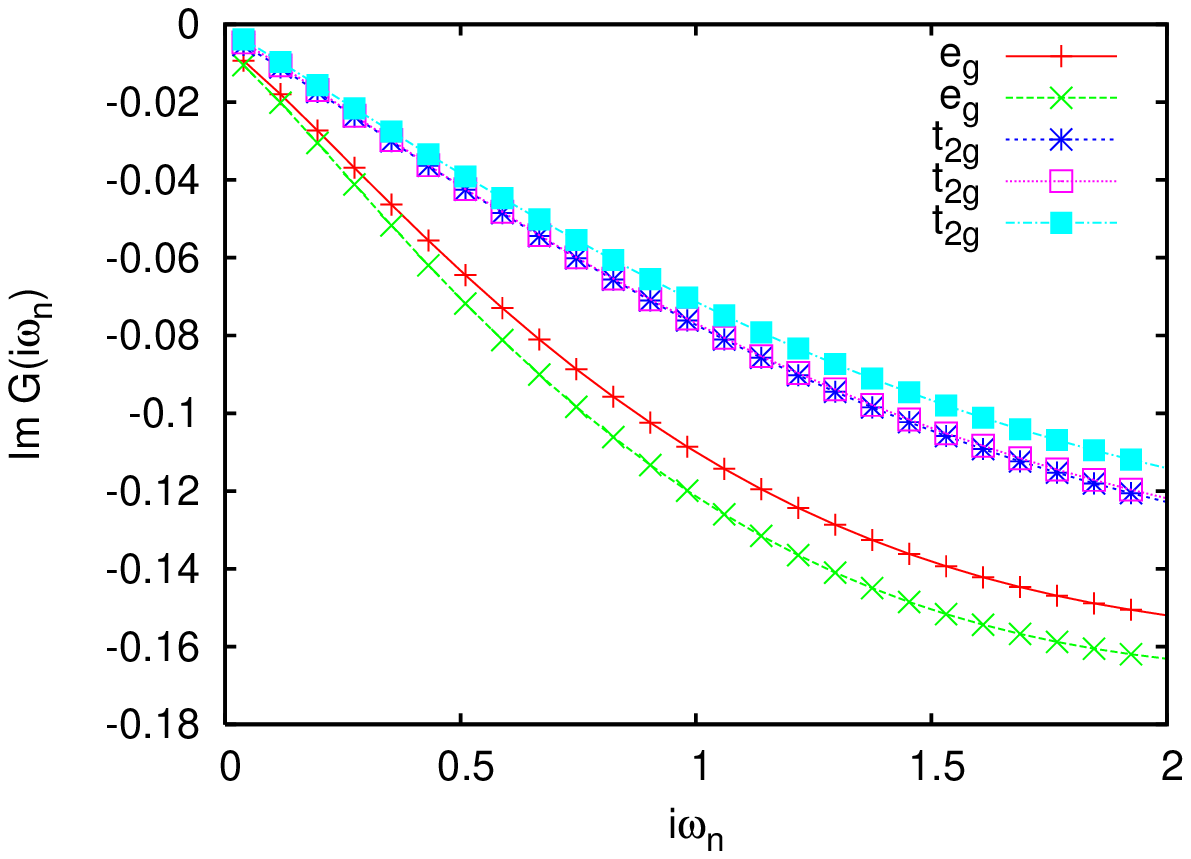}\\
  \caption{(Top) The imaginary part of the Green's function on the
    Matsubara frequencies of cubic \pmo. (Bottom) The imaginary part
    of the Green's function on the Matsubara frequencies of
    orthorhombic \pmo.}
  \label{fig:pmo-imgiw}
\end{figure}

The LDA band structures for the orthorhombic and cubic PMO is shown in
Fig.~\ref{fig:pmo-bnd}. It can be seen that the orthorhombic
distortions reduce the band width of the whole Mn $3d$ manifold (by
$\sim$0.7 eV) as well as of the $e_g$ subset of bands (by $\sim$1.0
eV) relative to that of the cubic structure. As a result of the
reduced band width, the on-site Coulomb repulsion and Hund's coupling
are effective in making the distorted structure insulating. This can
be seen from the opening of a gap in the orthorhombic structure in the
spectral functions of the Mn $3d$ orbitals as obtained from our
LDA+DMFT calculations that are presented in Fig.~4 of the main
text. We also present in Fig.~\ref{fig:pmo-imgiw} the imaginary part
of the Green's functions as a function of the Matsubara frequencies
for Mn $3d$ orbitals that were analytically continued using maximum
entropy method to obtain the spectral functions on the real axis. For
the cubic \pmo, the imaginary part of the Green's function for $e_g$
orbital extrapolates to a non-zero value at $\imath \omega_n = 0$,
indicating a finite density of states at the Fermi level. On the other
hand, the imaginary part of the Green's function for all orbitals
extrapolate to zero for the orthorhombic \pmo, indicating an
insulating state. It can also be seen that there is minimal noise in
the imaginary part of the Green's functions. The spikes in
the spectral function on the real axis apparent on Fig.~4 of the main
text are not due to numerical noise, but to sharp excitations involving multiplets of 
the Pr-$4f$ shell (treated here in the Hubbard-I approximation).

% \bibliography{refs_supp}

%\begin{references}
%
%  \bibitem{rini07} M. Rini, R. Tobey, N. Dean, J. Itatani, Y. Tomioka,
%    Y. Tokura, R. W. Schoenlein, and A. Cavalleri, Nature {\bf 449},
%    72 (2007).
%    
%  \bibitem{foer11} M. F\"orst, C. Manzoni, S. Kaiser, Y. Tomioka,
%    Y. Tokura, R. Merlin, and A. Cavalleri, Nature Phys. {\bf 7}, 854
%    (2011).
%  
%  \bibitem{kais13} S. Kaiser, D. Nicoletti, C. R. Hunt, W. Hu,
%    I. Gierz, H. Y. Liu, M. Le Tacon, T. Loew, D. Haug, B. Keimer, and
%    A. Cavalleri, arXiv:1205.4661 (preprint).
%    
%  % LaMnO3 structure
%  \bibitem{huan97} Q. Huang, A. Santoro, J. W. Lynn, R. W. Erwin,
%    J. A. Borchers, J. L. Peng, and R. L. Greene, Phys. Rev. B {\bf 55},
%    14987 (1997).
%
%  \bibitem{pbe} J. P. Perdew, K. Burke, and M. Ernzerhof,
%    Phys. Rev. Lett. {\bf 77}, 3865 (1996).
%
%  \bibitem{tobe08} R. I. Tobey, D. Prabhakaran, A. T. Boothroyd, and
%    A. Cavalleri, Phys. Rev. Lett. {\bf 101}, 197404 (2008).
%
%\end{references}

\end{document}